\documentclass[twocolumn]{aastex631}
\usepackage{amsmath}

\def\Msun{$M_{\rm \odot}$}

\def\ps{$\rm km \,s^{-1}\,kpc^{-1}$}
\def\arcdeg{$^\circ$}

\def\H2{$\rm H_2$}

\def\EffPotL1{$\Phi_{\rm eff,b}(R_{\rm L1/L2})$}

\begin{document}

\title{Solar System Migration Points to a Renewed Concept: Galactic Habitable Orbits}

\author[0000-0002-2154-8740]{Junichi Baba}
\affiliation{Amanogawa Galaxy Astronomy Research Center, Kagoshima University, 1--21--35 Korimoto, Kagoshima 890-0065, Japan.}
\affiliation{National Astronomical Observatory of Japan, Mitaka, Tokyo 181-8588, Japan.}

\author[0000-0002-9397-3658]{Takuji Tsujimoto}
\affiliation{National Astronomical Observatory of Japan, Mitaka, Tokyo 181-8588, Japan.}

\author[0000-0001-8226-4592]{Takayuki R. Saitoh}
\affiliation{Department of Planetology, Graduate School of Science, Kobe University 1-1, Rokkodai-cho, Nada-ku, Kobe, Hyogo, 657-8501, Japan.}

\begin{abstract}
Astrophysical evidence suggests that the Sun was born near 5 kpc from the Galactic center, within the corotation radius of the Galactic bar, around 6--7 kpc. This presents challenges for outward migration due to the Jacobi energy constraint, preventing stars from easily overcoming the corotation barrier. In this study, we use test particle simulations to explore two possible migration pathways for the Sun: a ``trapped'' scenario, where the Sun’s orbit was influenced by a slowing Galactic bar, and an ``untrapped'' scenario driven by dynamic spiral arms. Our results demonstrate that both mechanisms can explain how the Sun migrated from its birth radius ($\approx 5$ kpc) to its current orbital radius around 8.5--9 kpc. Furthermore, we investigate the environmental changes experienced by the Sun along these migration pathways, focusing on variations in radiation hazards and comet fluxes, which may have impacted planetary habitability. These findings highlight the dynamic nature of galactic habitability, emphasizing that the path a star takes within the Milky Way can significantly affect its surrounding environment and the potential for life. We propose a new concept of ``Galactic habitable orbits,'' which accounts for evolving galactic structures and their effects on stellar and planetary systems. This work contributes to a deeper understanding of the solar system’s migration and its implications for habitability within the Milky Way.
\end{abstract}


\section{Introduction}

\begin{figure}
\begin{center}
\includegraphics[width=0.5\textwidth]{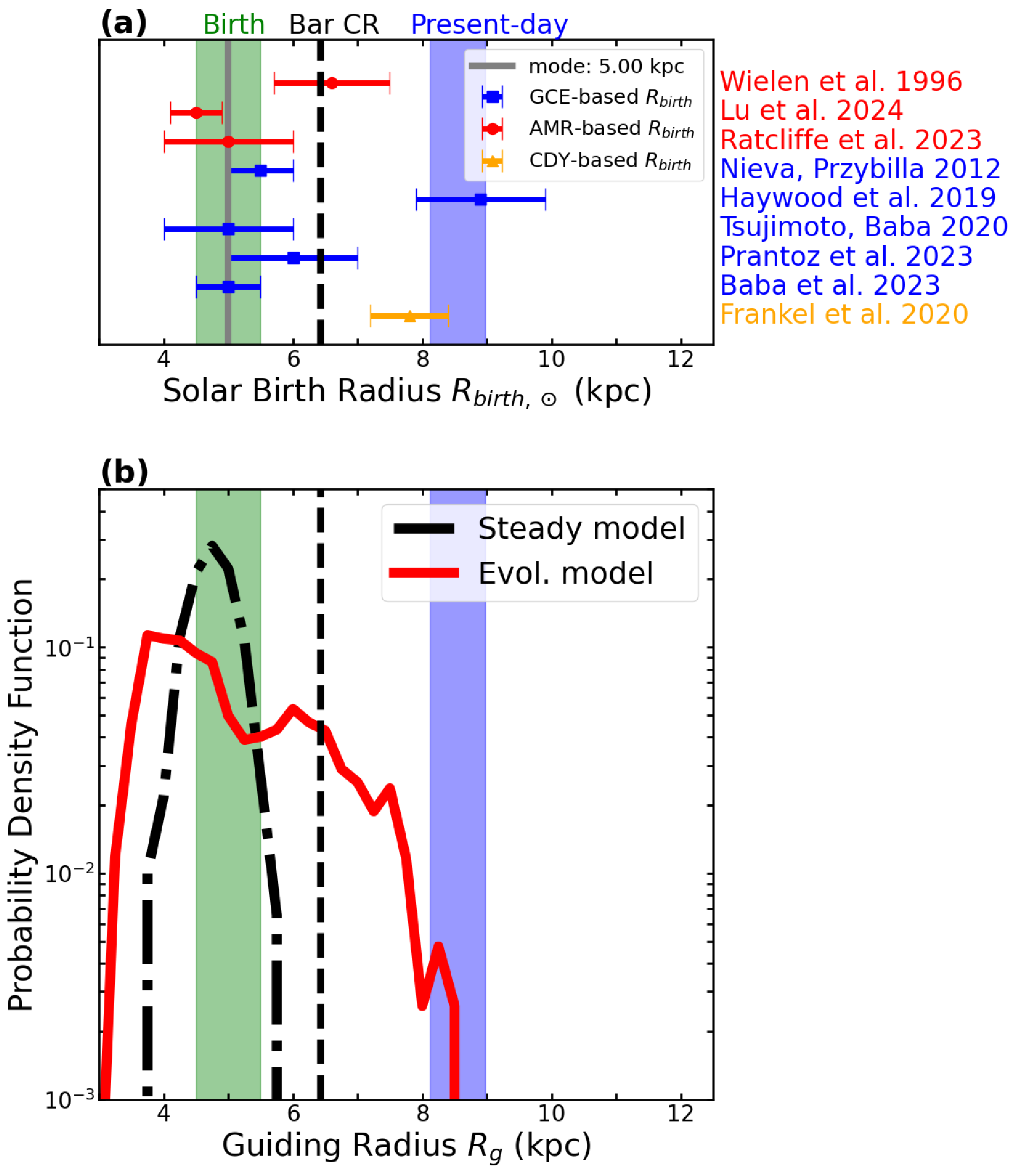}
\caption{
    Estimations of the solar birth radius ($R_{\rm birth,\odot}$) and probability density functions of eventual guiding radius ($R_{\rm g}$).
    Panel (a) Recent estimates of $R_{\rm birth}$ from Galactic chemical evolution models (GCE; blue squares)~\citep[][]{NievaPrzybilla2012,Haywood+2019,TsujimotoBaba2020,Prantzos+2023,Baba+2023}, the stellar age-metallicity relation (AMR; red dots)~\citep[][]{Wielen+1996,Lu+2024,Ratcliffe+2023}, and a chemodynamical model using an orbital diffusion approximation (CDY; orange triangle)~\citep[][]{Frankel+2020}, overlaid for comparison. 
    In cases where the references did not specify the range of $R_{\rm birth,\odot}$ estimates, an uncertainty of $\pm 1$ kpc was applied.
    The mode value of these recent estimations is indicated by a vertical solid line. The green-shaded region around 5 kpc denotes the adopted range of the Sun's birth radius for this study, while the blue-shaded region around 8.5 kpc indicates the present-day guiding radius of the Sun ($R_{\rm g,\odot}$). The vertical dashed line marks the present-day CR radius of the Galactic bar at 6.43 kpc.
    Panel (b) Eventual $R_{\rm g}$ probability density distributions of stars for different scenarios. 
    The solid dot-dashed line represents the ``steady model'' with a constant rotational speed for the bar and spiral arms ($\Omega_{\rm s}/\Omega_0=1.0$), with a spiral amplitude ratio of $\Sigma_{\rm s}/\Sigma_{\rm disk}=25$\%. 
    The red solid line depicts the ``evolving model'', which includes a decelerating bar and dynamic spiral arms, also at a spiral amplitude of $\Sigma_{\rm s}/\Sigma_{\rm disk}=25$\%. 
    Details on variations in the pattern speed and amplitude of the spiral arms are further examined in Figure \ref{fig:MigrationComp}.
}	
\label{fig:MigrationEffieincy}
\end{center}
\end{figure}

The formation and migration of the solar system within the Milky Way are crucial topics in astrophysics. The Sun's metallicity, [Fe/H], which is higher than that of nearby stars of similar age, suggests that it formed closer to the Galactic center. Estimates of the Sun's birth radius ($R_{\rm birth,\odot}$), derived from its elemental abundance and the metallicity gradient of the Galactic disk 4.6 Gyr ago, place its formation between 4 and 9 kpc from the Galactic center~\citep[][]{NievaPrzybilla2012,Haywood+2019,Frankel+2020,Prantzos+2023,TsujimotoBaba2020,Lu+2024,Ratcliffe+2023,Baba+2023}, with an average around 5 kpc (Fig.~\ref{fig:MigrationEffieincy}a). These estimates indicate that the Sun originated in an active region of the Milky Way. Its subsequent migration to its present-day position, approximately 8.2 kpc from the Galactic center~\citep[][]{Bland-HawthornGerhard2016}, highlights the dynamic influence of the Galactic bar and spiral arms~\citep[][]{SellwoodBinney2002,Kubryk+2013}. 
The Galactic bar is estimated to be 6--8 Gyr old~\citep[][]{Sanders+2024}, implying that it was present when the Sun was formed. With a corotation (CR) radius of around 6 to 6.5 kpc and a pattern speed ($\Omega_{\rm b}$) of 35--40 \ps{}~\citep[e.g.][]{ClarkeGerhard2022}, 
the presence of the bar suggests the Sun originated within this radius.

The Sun's initial placement within the bar's CR radius posed significant challenges for its outward migration because of Jacobi energy constraints~\citep[][]{BinneyTremaine2008}. Tracing the Sun's migratory path requires understanding how these constraints were overcome. The Jacobi energy in the rotating frame of the bar, defined as $E_{\rm J} = \frac{1}{2}v^2 + \Phi_{\rm eff,b}$, where $\Phi_{\rm eff,b} = \Phi - \frac{1}{2}\Omega_{\rm b}^2R^2$, combines gravitational and centrifugal potentials. The effective potential $\Phi_{\rm eff,b}$ is a convex function with a peak at the bar's CR radius, referred to as the ``CR barrier''. In general, stars cannot exist in regions where the effective potential exceeds the star's Jacobi energy ($\Phi_{\rm eff,b} > E_J$), since $E_{\rm J} - \Phi_{\rm eff,b} = \frac{1}{2}v^2 \geq 0$. For the Sun, this principle implies that being born inside the bar's CR radius requires overcoming the CR barrier to migrate outward. However, considering the near-circular nature of the Sun's orbit today, its Jacobi energy is insufficient to surpass the bar's CR barrier. Therefore, alternative mechanisms must have facilitated the Sun's migration inside the present-day bar's CR radius ($\approx 6.4$ kpc) to its current orbital radius ($\gtrsim 8.5$ kpc).

The interaction between the Galactic bar and spiral arms introduces complexities in the conservation of Jacobi energy, providing potential pathways for the Sun and other stars to overcome the bar's CR barrier. Recent data from the {\it Gaia} satellite confirm both the deceleration of the Galactic bar~\citep[][]{Chiba+2021,ChibaSchoenrich2021} and the dynamic nature of its spiral arms~\citep[][]{Asano+2024,Funakoshi+2024}. These findings highlight the critical role of the bar’s deceleration~\citep[][]{Halle+2015,Chiba+2021} and the transient nature of spiral arms~\citep[][]{SellwoodBinney2002,Grand+2012a} in stellar migration.

The orbital migration of the solar system, influenced by the evolving structure of the Milky Way, provides valuable insights into the Galactic habitable zone (GHZ)—a critical framework for assessing the potential for life within the Milky Way. The GHZ is generally defined as a region that balances sufficient heavy elements for planet formation with a lower occurrence of hazardous events like supernovae \citep[][]{Lineweaver+2004,Gowanlock+2011}. Expanding on this concept, \citet{KokaiaDavies2019} showed that stars closer to the Galactic center or plane encounter giant molecular clouds (GMCs) more frequently, which increases their exposure to nearby supernovae that could destabilize planetary orbits and alter atmospheres. Similarly, \citet{Spinelli+2021} demonstrated that the threat of gamma-ray bursts (GRBs) varies across the Milky Way, highlighting uneven GRB-related risks. \citet{Baba+2023} further modeled the Galactic chemical evolution of elements essential for planet formation (e.g., C, O, Mg, Si, Fe), providing insights into how their distribution may influence planetary system diversity and habitability across the Milky Way.

Building on these recent advances, our study investigates the solar system's orbital migration within the dynamically evolving galactic structures. By analyzing how the solar system’s position has changed over time due to interactions with the Galactic bar and spiral arms, we aim to refine our understanding of the GHZ and explore how such migrations could influence habitability on Galactic timescales.

\section{Migration of the Sun}

To investigate the migration of the solar system through the Milky Way's dynamically changing bar and spiral arms, we employ a test-particle approach to represent the time-varying potentials of these galactic structures. The galaxy model and orbital calculation methods follow \citet{TsujimotoBaba2020}, with some modifications to the model parameters (Appendix A). 
For the bar structure, we use the analytical model from \citet{Binney2018}, considering two scenarios: in one, the bar rotates at a constant pattern speed ($\Omega_{\rm b}$) of 36 \ps{} (rigid bar); in the other, the bar slows down from an initial speed of 55 \ps{} at the time of the solar system's formation to 36 \ps{} over 4.6 Gyr. This model reflects the gradual slowing of the bar's spin, with the CR radius being around 4 kpc at the time of the Sun's formation.
The spiral arms are modeled in two ways: one as a steady rotating pattern ~\citep[steady density wave, SDW;][]{Shu2016} and the other as a dynamic, winding pattern (DYN). The DYN arms are short-lived and reappear frequently, as seen in many $N$-body simulations~\citep[][]{Grand+2012a,Baba+2013}. These arms change in strength and shape over time due to the differential rotation of the galaxy, which causes the pattern speed ($\Omega_{\rm s} \approx V_{\rm c}/R$) and the pitch angle to vary. In contrast, the SDW model treats the spiral arms as long-lasting structures that rotate at a constant speed ($\Omega_{\rm s}$) and keep a fixed pitch angle ($i_{\rm s}$). Both types of arms are modeled using the analytic model of \citet{CoxGomez2002}, with DYN arms following \citet{Hunt+2018} to include time-varying pitch angles and amplitudes.

To explore the effect of the bar and spiral arm changes on solar migration, we compare four models: rigid bar + SDW (steady model), rigid bar + DYN, slowing bar + SDW, and slowing bar + DYN (evolving model).

\subsection{Migration Efficiencies}

The efficiency of solar migration is shown in Figure~\ref{fig:MigrationEffieincy}(b), which presents the final guiding radius ($R_{\rm g}$) distribution of the solar system in the simulations. These simulations started with 10,000 particles, initially placed in stable orbits within the bar potential, with birth radii ranging from $4.5 \leq R_{\rm g,birth} \leq 5.5$ kpc (green-shaded region). 
The dot-dashed line illustrates the outcome of the ``steady model,'' characterized by a pattern speed of $\Omega_s = \Omega_0$~\citep[][]{Barros+2021review}, where $\Omega_0$ represents the circular angular velocity of the Local Standard of Rest around the Galactic center, typically about 28.1 \ps{}~\citep[][]{Bland-HawthornGerhard2016}.
In this model, the fixed CR radius at 8.1 kpc and the inner Lindblad resonance at 1.1 kpc limit the potential for significant orbital migration, as these resonance radii are far from the Sun’s birth region~\citep[][]{Lynden-BellKalnajs1972}.

\begin{figure*}
\begin{center}
\includegraphics[width=0.95\textwidth]{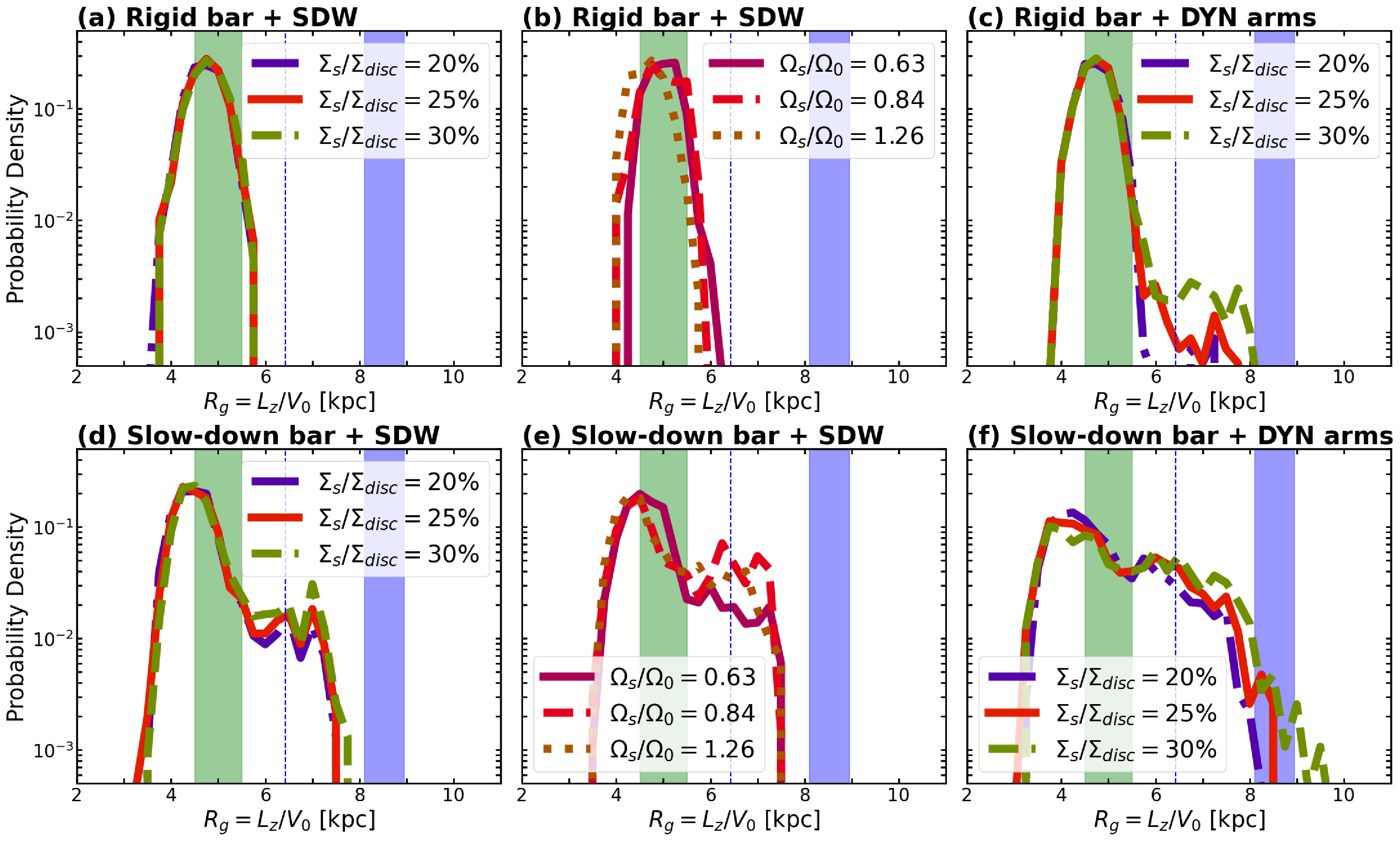}
\caption{
    Eventual $R_{\rm g} \equiv L_z/V_0$  distributions of stars with orbital eccentricity with $e<0.1$, which are initially set at $4.5 < R_{\rm g} < 5.5$ kpc (highlighted by the green shaded zone). 
    Panel (a) Cases with a rigid bar + SDW model. The spiral amplitudes are varied as $\Sigma_{\rm s}/\Sigma_{\rm disk}=$20\% (dot-dashed line), 25\% (solid line), and 30\% (dashed line). The vertical dashed line represents the radius of CR of the rigid bar at 6.43 kpc.
    Panel (b) Examines the effects of different pattern speed ratios ($\Omega_{\rm s}/\Omega_0$) in SDW models on the guiding radii distribution of stars. It presents cases with ratios of 0.63, 0.84, and 1.26, translating to $\Omega_{\rm s}$ values of 18, 24, and 36 \ps{}, respectively.
    Panel (c) investigates how combining a rigid bar with DYN arms influences stellar migration, with arm amplitudes of 20\% (dot-dashed line), 25\% (solid line), and 30\% (dashed line) showcasing varying migration efficiencies.
    Panels (d), (e), and (f) illustrate the migration outcomes with a slowing bar replacing the rigid bar, using the same spiral arm models as in Panel (a) to Panel (c), to compare how the bar's dynamics influence the Sun's migration.
}	
\label{fig:MigrationComp}
\end{center}
\end{figure*}

Figure~\ref{fig:MigrationComp}(a)–(c) compares the migration efficiency of the solar system under different spiral arm models in a rigid bar potential. Figures~\ref{fig:MigrationComp}(a) and \ref{fig:MigrationComp}(b) show that in the rigid bar + SDW model, varying the spiral amplitude $\Sigma_s/\Sigma_{\rm disk}$ or the pattern speeds $\Omega_s/\Omega_0$ of the SDW results in minimal orbital changes for the solar system. This indicates that altering these parameters has little effect on migration\footnote{
\citet{Martinez-Barbosa+2015} conducted similar dynamical studies using the rigidly rotating bar + SDW model and, consistent with our findings, observed minimal orbital migration for the Sun. Using backward-time integration to estimate the Sun's birth radius, these studies suggest that the Sun originated near its current position or possibly in the outer Galactic disk. This consistency with our results highlights the formidable barrier posed by the bar's CR, indicating that the modest increase in $E_{\rm J}$ provided by rigid spiral arms alone is insufficient to enable solar migration from within the bar's CR to its present location.
}. 
By contrast, Figure~\ref{fig:MigrationComp}(c) shows that the rigid bar + DYN arm model, with time-varying pitch angles, allows for much more effective solar migration, particularly when the spiral arm amplitude exceeds 30\%.

The influence of a slowing bar on migration is highlighted in Figures~\ref{fig:MigrationComp}(d)-\ref{fig:MigrationComp}(f). These figures show that combining a slowing bar with SDW models significantly enhances migration efficiency. However, when using a slowing bar alone, the solar system migration is confined to the vicinity of the current CR radius of the bar (6.5 kpc). While the slowing bar facilitates outward migration by capturing the solar system in its CR, the system remains trapped near the 6.5 kpc radius in the SDW model, preventing further outward migration.

The solid red line in Figure~\ref{fig:MigrationEffieincy}(b) represents the results from the fully ``evolving model,'' which incorporates both a slowing bar and dynamic spiral arms with a spiral amplitude of 25\%, identical to that of the steady model. After 4.6 Gyr, these particles spread over a much wider radial range, from $3 \lesssim R_g \lesssim 8.5$ kpc. Notably, about 0.8\% of these particles managed to migrate to the present-day Sun’s guiding radius range (blue-shaded region). Figure~\ref{fig:MigrationComp}(f) further demonstrates that even with a DYN arm amplitude of 20\%, the fully evolving model facilitates the solar system’s migration to its present-day radius. This success is due to the dynamic arm’s ability to vary its pitch angle, which allows for broader and more effective migration across the galactic disk. Therefore, dynamic changes in both the bar and spiral arms are essential for the efficient solar migration. Considering observational constraints on the amplitude of dynamic spiral arms (20--30\%), the migration range varies between 0.4\% and 1.4\%.

\subsection{Trapped and Untrapped Migration Pathways}

\begin{figure*}
\begin{center}
\includegraphics[width=0.95\textwidth]{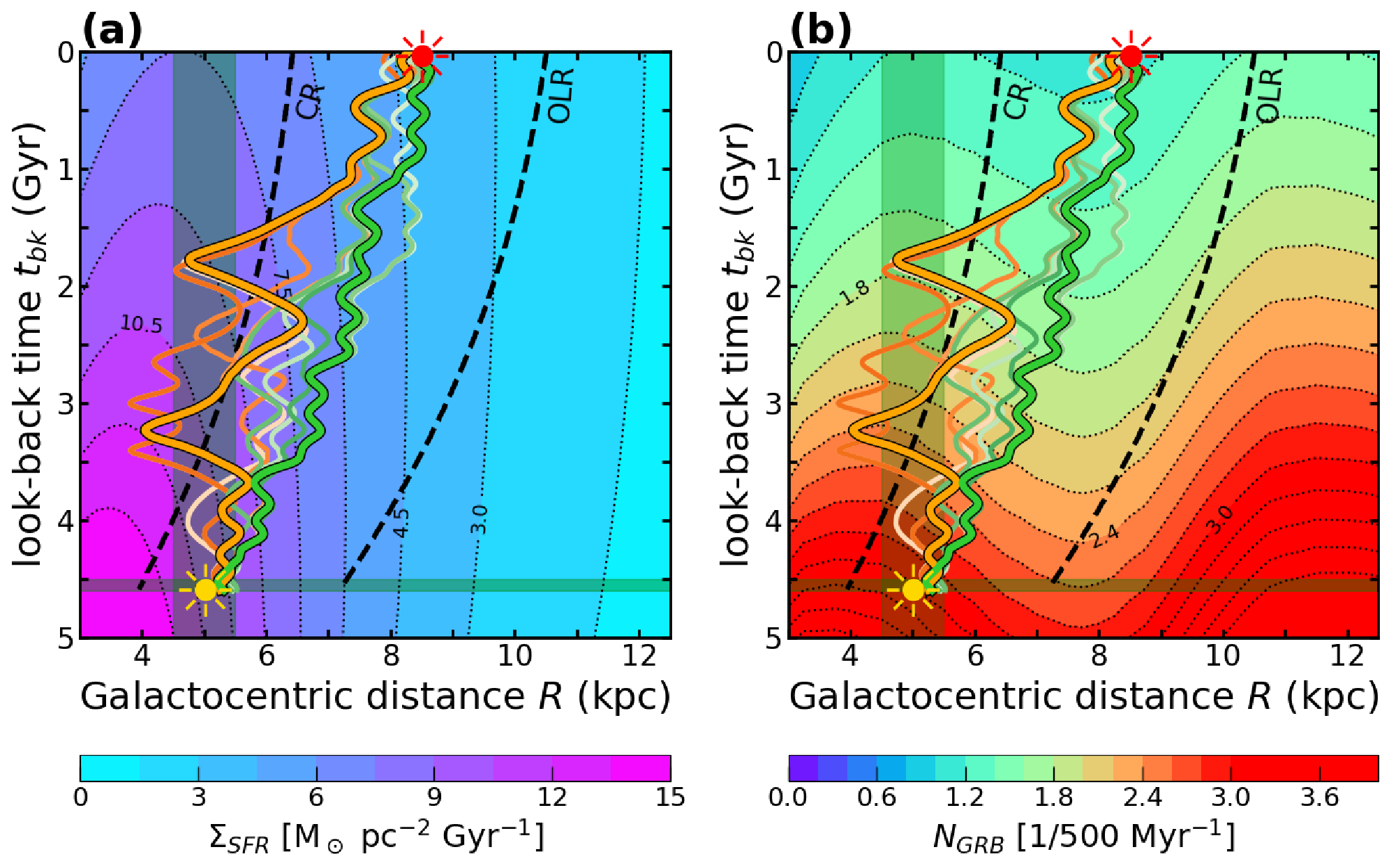}
\caption{
    Trajectories of the Sun on the $R$-$t_{\rm bk}$ planes for the evolving model.
    The orange lines show the trajectories of ``trapped migrators,'' while the green lines depict ``untrapped migrators.'' Thick lines represent the trajectories of Sun-like particles, while thin lines show orbits similar to those indicated by the thick lines. The background colors with contours indicate (a) the SFR density ($\Sigma_{\rm SFR}$) and (b) lethal GRB event rates ($N_{\rm GBR}$), based on the Galactic chemical evolution model \citep[][]{Baba+2023}. The black dashed lines indicate the locations of the bar's CR and OLR radii.
}	
\label{fig:MigrationGHZ}
\end{center}
\end{figure*}

To understand the migration pathways of the solar system, we examined the changes in the orbital radii of representative particles in the evolving model. 
Figure~\ref{fig:MigrationGHZ} illustrates the trajectories of the Sun-like particles, plotted on the $R$--$t_{\rm bk}$ plane, where $R$ represents the Galactocentric radius and $t_{\rm bk}$ (look-back time) measures the time elapsed from a specified past event to the present. The black dashed lines indicate the positions of the CR and the outer Lindblad resonance (OLR) of the bar, which are crucial for understanding the changes in stellar orbits due to orbital resonance~\citep[][]{Lynden-BellKalnajs1972}. We identified two types of migration paths: particles trapped by the bar's CR (orange) and those not trapped (green), which we refer to as ``trapped migrators'' and ``untrapped migrators,'' respectively.

Tracing the solar system's migration path within the Milky Way is challenging due to the complex nature of stellar orbital changes. However, studying solar siblings—stars that may have formed in the same stellar cluster as the Sun \citep[e.g.][]{Adams2010}—provides valuable insights into this history. These stars share similar chemical compositions and initial conditions, illuminating possible routes the solar system could have taken. As suggested by the migration probability of the evolving model (solid red line in Figure~\ref{fig:MigrationEffieincy}b), many particles are captured within the bar's CR around 6.5 kpc. By analyzing their distribution in phase space, we can infer whether these stars remained within the Galactic bar's CR radius, indicating trapped migration, or moved beyond it, suggesting untrapped migration. This analysis helps us to understand the Sun's migratory path more clearly.

\subsection{Solar Migration in a Fast, Short Bar Case}

Recent studies suggest that the Galactic bar may be shorter and faster than assumed in our study, with a pattern speed around $\Omega_b \approx 55$ \ps{}~\citep[e.g.,][]{Vislosky+2024}, shifting the CR radius to about 4 kpc. If, as indicated in the literature (Figure \ref{fig:MigrationEffieincy}a), the solar system formed around 5 kpc, it would have been positioned outside the CR radius of this shorter, faster bar. Additionally, dynamical friction with dark matter \citep[][]{Chiba+2021} implies that the bar would have been even shorter and faster at the time of the solar system's formation. Together, these factors suggest a minimal influence of the bar on early solar migration, making it more likely that migration was driven primarily by interactions with spiral arms.

If this was the case, solar migration would likely have been influenced more by DYN arms, as SDWs typically induce minimal migration. Thus, with a fast, short bar and dynamic spiral arms, the solar system likely followed an ``untrapped migration'' pathway.

\section{Changes of Surrounding Environments along Migration Pathways}

The migration of the solar system through the Milky Way has not only altered its position in the galaxy but also led to significant changes in its surrounding environments over time. These environmental changes have had critical implications for planetary habitability \citep[e.g.][]{Ruderman1974} and the supply of essential life materials \citep[e.g.][]{Watson+2024}. In the following sections, we explore how the varying radiation hazards and comet fluxes influenced by the Sun's migration pathways have shaped the conditions for life in the solar system.

\subsection{Radiation Hazard Variations}
\label{sec:radiation}

We examine how the solar system's migration through the Milky Way has altered radiation hazards, focusing specifically on the star formation rate (SFR) density and GRB event rates, both of which significantly influence planetary habitability. 
High SFRs are associated with frequent supernovae, as massive stars rapidly reach the end of their lifetimes. These supernovae can substantially impact their surrounding environments, especially through lethal GRBs. GRBs are divided into two types: short-duration GRBs (SGRBs), originating from compact object mergers~\citep[][]{Berger2014} and common in older stellar populations, and long-duration GRBs (LGRBs), resulting from massive star collapses~\citep[][]{WoosleyBloom2006} in star-forming regions. Both types of GRBs pose significant risks to life by exposing planets to intense high-energy radiation.

To quantify these hazards, we derived the SFR density from our Galactic chemical evolution model~\citep[][]{Baba+2023} and evaluated the lethal GRB rate, defining lethal radiation as an energy flux $F \geq F_c = 10^8~{\rm erg~cm^{-2}}$ within a critical distance $d_c$ of the GRB (see Appendix \ref{sec:evalGRB} for details), following \citet{Spinelli+2021}. Such an energy flux can produce stratospheric nitrogen compounds (NO$_x$), rapidly destroying about 90\% of the ozone layer and exposing the surface to harmful levels of solar UVB radiation \citep[][]{Thomas+2005}.
The background colors and contours in Figure~\ref{fig:MigrationGHZ} illustrate the SFR density $\Sigma_{\rm SFR}$ (left) and the lethal GRB event rate $N_{\rm GRB}$ (right). Trapped migrators (orange lines) tend to remain in the Milky Way's inner regions, believed to be the Sun's birthplace 4.6 Gyr ago, marked by high SFRs and frequent supernova events. Figure~\ref{fig:MigrationGHZ}(a) shows that the Sun experienced higher $\Sigma_{\rm SFR}$ than the present solar neighborhood for up to 2 Gyr. Figure~\ref{fig:OrbitGHZ}(a) highlights the changes in SFR densities along the trapped (orange) and untrapped (green) migration. The inner disk, during the solar system's formation, was about 2.5 times richer in star formation activity. Trapped migration keeps the solar system in high-SFR regions with high supernova frequencies, whereas untrapped migration (green lines) leads to a steady increase in orbital radius and a decline in nearby star formation activities and supernova events.

The inner regions of the Milky Way, where the Sun was born, also had elevated lethal GRB rates, particularly from SGRBs, about three times higher than the present-day solar neighborhood (Fig.~\ref{fig:MigrationGHZ}b). Figure~\ref{fig:OrbitGHZ}(b) shows that the Sun probably experienced these elevated GRB rates for up to 2 Gyr after its formation. Although these regions are rich in elements essential for planet formation, they present significant risks to life as a result of intense high-energy radiation. These environmental factors, combined with frequent supernovae and GRBs, presented considerable challenges to the sustainability of emerging life.

The migration process has moved the solar system from hazardous inner regions to a relatively safer and more stable environment. ``Escaping'' the severe radiation hazards of the inner Galaxy, the solar system currently finds refuge in an area with lower frequencies of supernovae and other catastrophic events, potentially allowing life on Earth to develop and evolve. However, further migration to outer regions could expose the Sun to more severe radiation environments~\citep[][]{Spinelli+2021}, since LGRBs occur more frequently among low-metallicity massive stars~\citep[][]{Virgili+2011}. If the Sun continues to move further out into the Milky Way, driven by interactions with dynamic spiral arms, the solar system could face greater risks, including higher exposure to radiation and other hazardous conditions in the outer Galactic regions.

\subsection{Comet Flux Variations}
\label{sec:comets}

The migration path of the solar system significantly influences the comet flux entering the planetary region, which affects the supply of life-building materials. Galactic tidal forces~\citep[e.g.][]{HeislerTremaine1986} and nearby stellar encounters~\citep[e.g.][]{Rickman+2008} exert gravitational perturbations on the Oort cloud, influencing the influx of long-period comets. These comets are rich in prebiotic molecules such as hydrogen cyanide (HCN) and simple amino acids, as well as essential elements like phosphorus monoxide (PO), which are crucial for the delivery of prebiotic and organic materials~\citep[][]{Watson+2024}. Understanding changes in comet flux is thus critical for assessing the potential supply of these life-building materials.

The variation in comet flux caused by galactic tides is estimated by analyzing the time-varying effects of the galactic tidal field on the solar system as it orbits the Milky Way, following the method of \citet{Gardner+2011}. Since the vertical component of the galactic tidal force, $G_z$, dominates over the in-plane components within the galactic disk~\citep[][]{HeislerTremaine1986}, we consider only $G_z$ in our analysis.
Ignoring the contributions from the bar and spiral arms, this study evaluates $G_z$ and scales it according to \citet{Gardner+2011} to estimate the comet flux as $f_{\rm comets} \approx 10 \left(\frac{G_z}{4.5\times 10^3~{\rm (km~s^{-1}~kpc^{-1})^2}}\right)~{\rm comets~year^{-1}}$.

\begin{figure*}
\begin{center}
\includegraphics[width=0.95\textwidth]{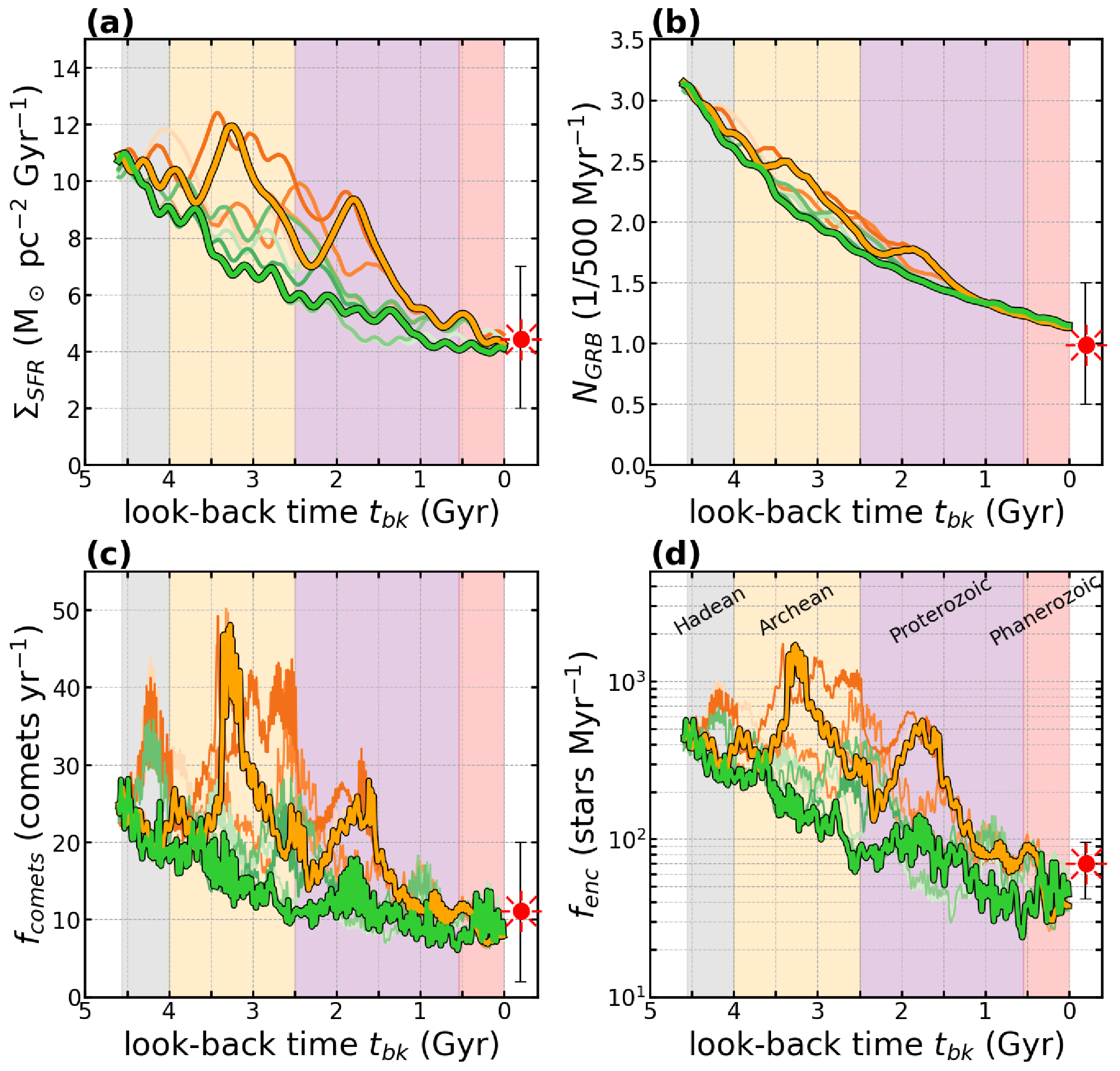}
\caption{
    Histories of Surrounding Environmental Changes Along the Potential Sun's Orbits.
    The environmental changes along the Sun's potential orbits are distinctly illustrated for ``trapped migrators'' (depicted in orange colors) and ``untrapped migrators'' (depicted in green colors). Panel (a) Shows star formation rate (SFR) densities, Panel (b) details lethal gamma-ray burst (GRB) event rates, Panel (c) depicts comet flux due to Galactic tides, and Panel (d) presents frequencies of stellar encounters. The error bars represent the estimated values in the current solar neighborhood taken from the literature~\citep[][]{Soler+2023,Rickman+2008,Thomas+2005,Bailer-Jones2018}.
    The vertical shaded regions in each panel indicate the geological eras of the Earth (Hadean, Archean, Proterozoic, and Phanerozoic). 
}	
\label{fig:OrbitGHZ}
\end{center}
\end{figure*}

The solar system's migration pathways significantly influence comet flux ($f_{\rm comets}$) variations over time. As shown in Figure~\ref{fig:OrbitGHZ}(c), in the solar system's birth environment, the comet flux was potentially about twice as high as it is currently, as suggested by previous studies~\citep[][]{Kaib+2011}. 
During trapped migration (orange lines), where the solar system remains within the inner disk for extended periods, the high comet flux periods are prolonged compared to untrapped migration (green lines). Trapped migrators oscillate due to being captured by the bar's CR, periodically moving closer to the Galactic center with a period of approximately 1 Gyr. This oscillatory motion could lead to experiencing significantly high comet flux, with the potential to encounter $f_{\rm comets}\approx 30$--$40~\rm comets~yr^{-1}$ twice during the first 2 Gyr.

The rate of stellar encounters is calculated based on the local density and velocity dispersion of stars along the Sun’s orbit using the formula $f_{\rm enc} = \pi D^2 n_\ast v_\ast$, where $D$ is the maximal encounter distance, $n_\ast$ is the stellar density (derived from our galaxy model), and $v_\ast$ is the relative velocity of stars to the Sun. 
A distance of $D=2$ pc is used to exclude distant encounters that are unlikely to affect the solar system~\citep[][]{Rickman+2008}. The velocity $v_\ast$ combines the Sun's motion and the velocity dispersion of nearby stars, which is derived from radial, azimuthal, and vertical components. The radial velocity dispersion $\sigma_R$ follows an exponential decay with a scale length of 2.5 kpc with $\sigma_{R} = 35$ km/s at $R=8$ kpc, and $\sigma_z$ is set at 10 km/s, reflecting typical vertical stellar motion in the Milky Way disk \citep[][]{Bland-HawthornGerhard2016}.

Nearby stellar encounters significantly impact comet flux variations. As indicated in Figure~\ref{fig:OrbitGHZ}(d), in the Sun's birth environment, the rate of nearby stellar encounters ($f_{\rm enc}$) was possibly about 10 times higher than it is today. Although galactic tidal fields determine a steady state of comet flux, stellar encounters can cause temporary increases, leading to comet showers~\citep[][]{Rickman+2008}. During trapped migration (orange lines), the solar system likely experienced a higher rate of these encounters, especially early in its history, potentially triggering frequent comet showers. These showers could have occurred twice during the first 2 Gyr, potentially influencing early Earth’s environment and the conditions for life. Frequent comet showers coinciding with the Archean era—when life is believed to have originated—suggest that such events may have contributed essential materials for the development of life. The high frequency of stellar encounters during this period raises the possibility that these comet showers played a key role in supplying Earth with life-building materials.

Our discussion has primarily focused on the effects of Galactic tides and nearby stellar encounters on the Oort cloud. However, encounters with GMCs, which are not included in the current study, may also play a significant role \citep{NapierStaniucha1982,HutTremaine1985}. Modeling the influence of GMCs is complex, as it requires assumptions about their mass, size, and lifetimes. Recent simulations further complicate this picture, showing that GMCs often have dynamic, filamentary structures that evolve due to galactic rotation and stellar feedback \citep[e.g.,][]{Fujimoto+2023,Baba+2017}. Given these complexities, \citet{KokaiaDavies2019} used test particle simulations to show that GMC encounter rates are higher near the Galactic center. Following their results, if, as our study suggests, the solar system originated around 5 kpc and later migrated outward, it would have encountered GMCs more frequently in its early history, likely causing gravitational perturbations that influenced the Oort cloud’s structure. As the solar system moved outward, the reduced frequency of GMC encounters may have created a more stable environment, potentially favorable for life.

\section{Galactic Habitable Orbits}

Our study shows that stars born at the same Galactocentric distance and currently existing at the same distance can experience vastly different environments for the habitability and evolution of planetary systems, depending on their orbital migration paths. These differences arise from variations in hazardous radiation environments and the supply of essential life materials encountered along their journeys. Therefore, galactic habitability is not solely determined by a star's present-day position from the Galactic center but is also strongly influenced by its migration history from its birthplace to its present-day location.

In other words, our findings suggest that galactic habitability is dynamic and heavily dependent on orbital migration processes rather than representing a fixed “zone,” as the traditional concept of the GHZ implies~\citep[][]{Lineweaver+2004,Gowanlock+2011,Spinelli+2021}. To link planetary and life evolution with galactic evolution, it is essential to consider the history of orbital migration rather than a simple region. We propose the concept of “Galactic habitable orbits,” which are pathways through the Milky Way offering varying conditions for life’s development based on evolving galactic dynamics. By considering the dynamical effects of the Galactic bar and spiral arms, we can better understand habitability in the Galactic context. Examining the differences in radiation environments and the supply of life-building materials encountered along different migration pathways provides a more nuanced understanding of how the dynamic nature of the Milky Way impacts planetary habitability.

Our approach to assessing habitability risks, such as lethal radiation exposure, relies heavily on ``Earth-based criteria,'' particularly the presence of an ozone layer that shields life from harmful UVB radiation \citep[e.g.,][]{Ruderman1974,Thomas+2005}. This raises the question of whether similar protective mechanisms are essential for life on other planets or if alternative atmospheric compositions might offer different forms of radiation resistance. For instance, if life can exist without an ozone layer, radiation hazards could impact habitability in ways that differ from our Galactic habitable orbits framework. As JWST’s advancements in observing exoplanetary atmospheres enable us to detect biosignature gases and analyze diverse atmospheric compositions, we gain new insights into habitability conditions that extend beyond Earth-centric models. Integrating these findings with models of Galactic chemical evolution and dynamical evolution will enrich our understanding of how varied Galactic environments may support life, emphasizing the need for a paradigm shift—from static zones to dynamic orbits—when evaluating the potential for life throughout the Milky Way.

\section*{Acknowledgments}
We are grateful to the anonymous referee for the constructive and insightful comments, which have significantly improved this manuscript. The authors acknowledge valuable discussions with Rimpei Chiba, Daisuke Kawata, and Yusuke Fujimoto regarding stellar migration driven by slowing bars, dynamic spiral arms, and giant molecular clouds. This research was supported by the Japan Society for the Promotion of Science (JSPS) under Grant Numbers 21K03633 and 21H00054. JB received additional support from the JSPS Grant Numbers 18K03711 and 24K07095. TT was supported by JSPS Grant Numbers 22K18280 and 23H00132. TRS acknowledges support from JSPS Grant Numbers 22H01259, 22K03688, and 21K03614.

\vspace{5mm}
\appendix

\section{Galaxy Models}
\label{sec:models}

To explore the impact of dynamically evolving bar and spiral arms on the Sun's orbital changes, we conducted orbital calculations of stars (test particles) under a time-dependent analytic gravitational potential. 
By controlling the parameters of the bar and spiral models, we can impose specific conditions that restrict and define the evolutionary scenarios within which the solar system could have migrated. This approach enables us to systematically explore how different configurations of the Milky Way's structural components influence the path and speed of migration, offering insights into the redistribution mechanisms of stars within the Milky Way galaxy. The simulations aim to isolate the effects of these dynamic structures on stellar orbits, thereby providing a deeper understanding of the galactic dynamics that govern the redistribution of stars across the Milky Way.


The total gravitational potential of the Milky Way galaxy is expressed as:
\begin{align}
 \Phi(R,\phi,z,t) = \Phi_{\rm 0}(R,z) + \Phi_{\rm bar}(R,\phi,z,t) + \Phi_{\rm spiral}(R,\phi,z,t),
\end{align}
where $\Phi_{\rm 0}$, $\Phi_{\rm bar}$ and $\Phi_{\rm spiral}$ represent the axisymmetric, bar, and spiral arm potentials, respectively; 
$(R,\phi,z)$ are the polar coordinates ($\phi$ is positive for clockwise direction), and the origin of the time ($t$) is the birth time of the Sun. The simulations were conducted for up to $t=t_0 \equiv 4.6$ Gyr.  The simulations presented here integrate $10^4$ particles with a fourth-order Runge-Kutta method. All particles are integrated simultaneously with a time step of 0.1 Myr.

We adopt the analytical gravitational potential model of the axisymmetric Milky Way model, $\Phi_{\rm 0}(R)$, which consists of a dark matter halo, a bulge, and thin+thick stellar disks. We assume a Navarro-Frenk-White halo with a virial mass of $9.5\times10^{11}$, a virial radius of $214$ kpc and a concentration parameter of $13.7$, and a Hernquist bulge with a mass of $4.3\times10^9$~\Msun{} and a scale length of $0.35$ kpc. Both stellar disks follow an exponential profile. We set the mass at $4.5\times10^{10}$\Msun{}, the scale length at $2.4$ kpc, and the scale height at $0.3$ kpc for the thin disk, and $8\times10^{9}$\Msun{}, $2.0$ kpc and $0.9$ kpc for the thick disk, respectively~\citep[][]{Bland-HawthornGerhard2016}. The parameters were chosen to fit the kinematic data of the Milky Way galaxy. This potential model sets the circular velocity ($V_{\rm c}$) at the solar radius $R_0 = 8.2$ kpc to be $V_0 = 230~\rm km~s^{-1}$ (Fig. \ref{fig:galmodel}a), which corresponds to the circular angular velocity $\Omega_0 = 28.1~\rm km~s^{-1}~kpc^{-1}$~\citep[][]{Bland-HawthornGerhard2016}.

\subsection{Bar Potential and Its Evolution}

\begin{figure*}
\begin{center}
\includegraphics[width=0.95\textwidth]{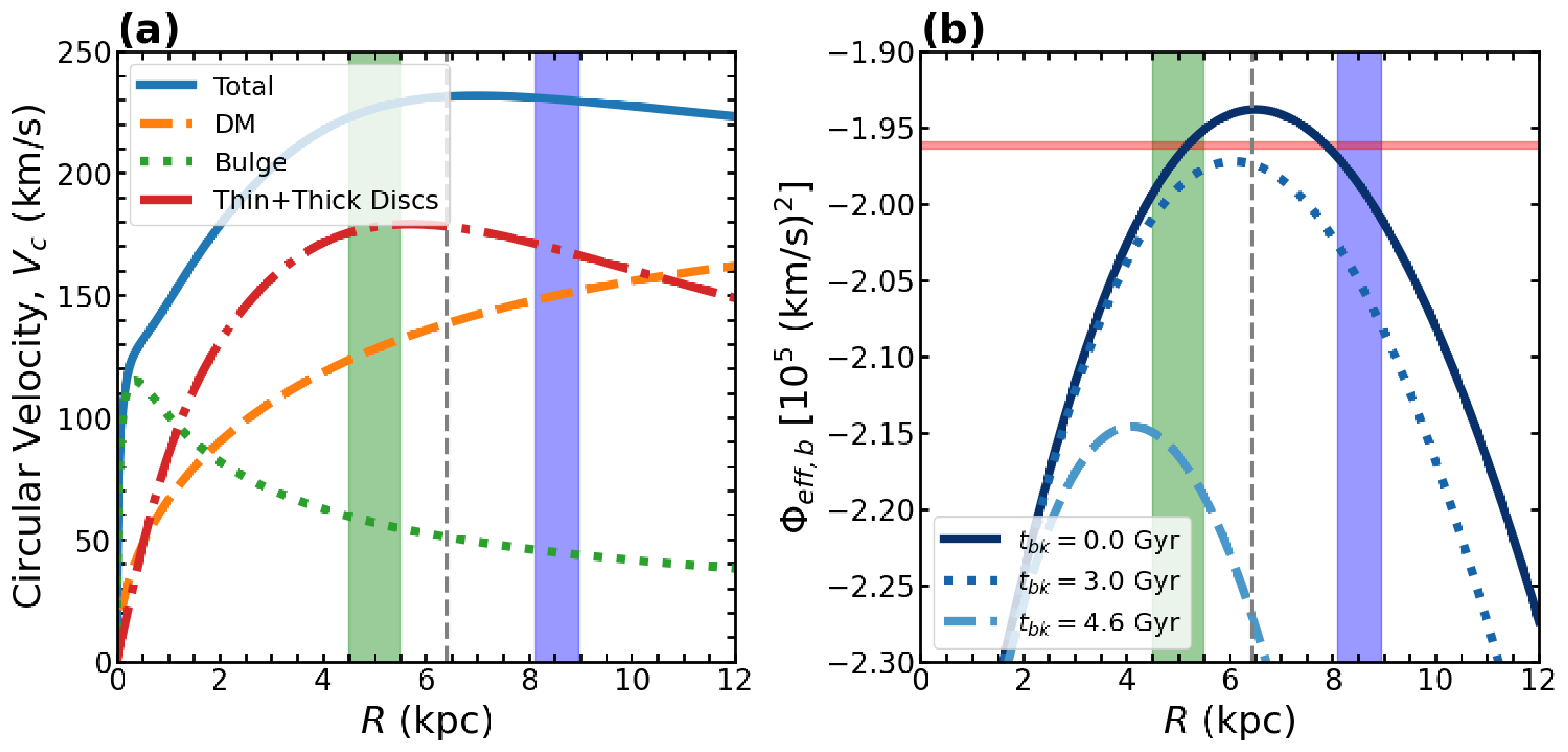}
\caption{
    Panel (a) Circular velocity ($V_{\rm c}$) and contributions of individual components as a function of the galactocentric radius, $R$. The green-shaded region around 5 kpc denotes the adopted range of the Sun's birth radius ($R_{\rm birth,\odot}$) for this study, while the blue-shaded region around 8.5 kpc indicates the present-day guiding radius of the Sun ($R_{\rm g,\odot}$). 
    The vertical dashed line marks the present-day CR radius of the Galactic bar at 6.43 kpc.
    Panel (b) The radial profiles of the effective potentials along the bar's major axis, $\Phi_{\rm eff,b}$, for these slow-down bar models are shown for look-back times $t_{\rm bk} = 0$, 3.0 and 4.6 Gyr ago. At each of these times, the peak of the profile corresponds to the CR radius of the bar, with respective values of 6.43 kpc, 6.1 kpc and 4.1 kpc. The horizontal red line indicates the current value of the Sun's Jacobi energy.
}	
\label{fig:galmodel}
\end{center}
\end{figure*}

To model the bar potential, following~\citep[][]{Binney2018}, we adopt a bar potential of the form
\begin{align}
\label{eq:barpot}
\Phi_{\mathrm{bar}}(R,\phi,z,t) = 
    -A_{\rm b}\frac{V_0^2 r_{\rm q}^3 R^2}{(R_{\rm b}^2+m^2)^{5/2}} 
    \cos[2(\phi - \phi_{\rm b} - \Omega_{\rm b} t)],
\end{align}
where $m^2 \equiv R^2 + z^2/q_{\rm b}^2$ and $r_{\rm q} = 1.5$ kpc and $q_{\rm b} = 0.9$; and $\phi_{\rm b}$ is the viewing angle of the major axis of the bar at the final state, fixed at $25^\circ$. $R_{\rm b}$ and $\Omega_{\rm b}$ are the scale length of the bar and the pattern speed of the bar, respectively, with $A_{\rm b}$ being the normalization of the bar strength. Considering recent studies suggesting the Galactic bar's age as between 6 and 8 Gyr~\citep[][]{Sanders+2024}, we infer that the bar exists before the Sun's birth. For this study, we set $A_{\rm b}=0.4$ as a standard value in our simulations.

In this study, we consider two types of bar model: a ``rigidly rotating'' bar and a ``slowdown'' bar. For the first model, we assume that the Galactic bar remains steadily rotating over time with a constant pattern speed, set at $\Omega_{\rm b,0}=36$ \ps{}~\citep[][]{Binney2018,ChibaSchoenrich2021}, in line with recent estimates~\citep[][]{ClarkeGerhard2022}. The second bar model posits a large initial pattern speed that decreases with time, reflecting $N$-body simulation predictions that stellar bars slow down due to dynamical interactions with dark halo and disk~\citep[e.g.][]{DebattistaSellwood1998}. To model this, we use an exponential slowdown function:
\begin{align}
 \label{eq:baromega}
 \Omega_{\rm b}(t) = \frac{\Omega_{\rm b,SSF}-\Omega_{\rm b,0}}{1-e^{-t_0/t_{\rm slow}}}(e^{-t/t_{\rm slow}}-1) + \Omega_{\rm b,SSF}.
\end{align}
Here $t_{\rm slow}$ is a slowdown timescale, with the pattern speed $\Omega_{\rm b,SSF}$ at $t = 0$ (i.e. the Sun's birth time) set at 55 \ps{}. Considering the significant uncertainty in the initial pattern speeds during bar formation for a galaxy similar to the Milky Way, we selected $t_{\rm slow}= 2$ Gyr as a standard value.

The length of the bar is related to its CR radius as indicated by theoretical and observational studies~\citep[][]{ContopoulosGrosbol1989}.
To incorporate this relationship, we model the bar length as
\begin{align}
 \label{eq:barlength}
 R_{\rm b}(t) = 2.09~{\rm kpc}~\times \frac{R_{\rm CR}(t)}{R_{\rm CR}(t_0)},
\end{align}
with $R_{\rm b} = 2.09$ kpc at $t = t_0$~\citep[][]{Binney2018}.
Figure~\ref{fig:galmodel}(b) depicts the time evolution of the effective potential, $\Phi_{\rm eff,b}(R) = \Phi(R) - \frac{1}{2}\Omega_{\rm b}^2R^2$, of the slowing bar model. The effective potential peak is lower during periods of faster pattern speed and rises and shifts outward as the bar slows.

\subsection{Spiral Potential and Its Evolution}

For our spiral potential model, we adopt the analytic expression of the logarithmic spiral~\citep[][]{CoxGomez2002}, 
\begin{align}
 \Phi_{\rm spiral}(R,\phi,z) =
    -4\pi G h_{\rm s}\rho_{\rm s}A_{\rm s}(t)
    \exp\left(\frac{R_0-R}{R_{\rm s}}\right) 
    \times \sum_{n=1,2,3}\frac{C_n}{K_nD_n}\cos(n\gamma_{\rm s})\left[{\rm sech}\left(\frac{K_nz}{\beta_n}\right)\right]^{\beta_n},
\end{align}
where
\begin{align}
 K_n = \frac{nm_{\rm s}}{R\sin i_{\rm s}},\\
 \beta_n = K_n h_{\rm s} (1+0.4K_n h_{\rm s}),\\
 D_n = \frac{1+K_nh_{\rm s}+0.3(K_nh_{\rm s})^2}{1+0.3K_nh_{\rm s}},\\
 \gamma_{\rm s} = m_{\rm s}\left(\phi+\Omega_{\rm s} t - \frac{\ln(R/R_0)}{\tan i_{\rm s}}\right).
\end{align}
Here, $m_{\rm s}$ is the number of the spiral arms, $i_{\rm s}$ is the pitch angle, and $\rho_{\rm s}$ is the density at $R_0$, $R_{\rm s}$ is the radial scale length of the spiral arm, and $h_{\rm s}$ is the scale height of the arm. $C_n$ is constant to control the azimuthal profile of the arms. 
Following \citet{CoxGomez2002}, we set $C_1 = 8/3\pi$, $C_2 = 1/2$, and $C_3 = 8/15\pi$, resulting in an azimuthal profile that behaves approximately as a squared cosine in the arms and is flat in the interarm region. 
$A_{\rm s}(t)$ is the temporal function to control the strength of the spiral arms (see below).
In general, the spiral arms disappear within the bar region. Thus, we damp the spiral potential in $R\lesssim 2R_{\rm b}(t)$, which is approximately 4 kpc at $t_{\rm bk}=0$~\citep[][]{Bland-HawthornGerhard2016}.
Considering the ongoing debate over whether the Milky Way galaxy has two or four arms, we focus only on $m_{\rm s}=2$ in this study. 

We consider two types of spiral arm models: the ``dynamic spiral'' (DYN) and classical ``(quasi-)stationary density wave'' (SDW) models. 
The DYN arms are transient and recurrent patterns, as seen in many $N$-body simulations of galactic disks, which not only exhibit amplitude variations over time but also evolve due to the differential rotation of the galactic disk~\citep[][]{Grand+2012a,Baba+2013}.
This leads to a nonconstant pattern speed, with $\Omega_{\rm s} \approx V_{\rm c}/R$, resulting in a time-dependent pitch angle. 
On the other hand, the SDW model represents a more traditional view, where spiral arms are long-lived, quasi-stationary structures, rotating at a constant speed, $\Omega_{\rm s}$, and maintaining a fixed pitch angle, $i_{\rm s}$~\citep[][]{Shu2016}.

To model the DYN arms, we adopt the ``winding'' spiral potential proposed by \citet{Hunt+2018}:
\begin{align}
 \Omega_{\rm s}(R) = \frac{V_{\rm c}(R)}{R},\\
 A_{\rm s}(t) = \exp\left(-\frac{(t-t_{\rm peak})^2}{2\sigma_{\rm s}^2}\right),
\end{align}
where $t_{\rm peak}$ is the time when the potential reaches the maximum amplitude, and $\sigma_{\rm s}$ is the characteristic timescale of the spiral lifetime. In our model, we set $\sigma_{\rm s} (t<t_{\rm peak}) = 0.5 \sigma_{\rm s}(t>t_{\rm peak})$ to model the $i_{\rm s}$--$A_{\rm s}$ relation predicted by dynamic spirals emerging in self-consistent $N$-body simulations, where the spiral arm is weak in the leading phase ($i_{\rm s}>90$\arcdeg{}) and reaches the maximum amplitude around $i_{\rm s} \approx 20$--$30$\arcdeg{}~\citep[][]{Baba+2013,Baba2015c}. In this study, we set $\sigma_{\rm s}(t>t_{\rm peak}) = 200$ Myr, which is a characteristic timescale of dynamic spirals in $N$-body simulations~\citep[][]{Baba2015c,Fujii+2018}.

To model the SDW models, we fix the speed of the spiral pattern, $\Omega_{\rm s}$, the pitch angle, $i_{\rm s}$, and the amplitude ($A_{\rm s}=1$). \citet{Eilers+2020} suggested a value of $\Omega_{\rm s} = 12$ \ps{}, or $0.42\Omega_0$ to reproduce the observed mean Galactocentric radial velocities of the red giant stars in the Milky Way, while \citet{Barros+2021review} suggested that the Sun lies near the corotation circle of the spiral arms, i.e. $\Omega_{\rm s} \approx \Omega_0 = 28.1$ \ps{}. 
We set $\Omega_{\rm s}/\Omega_0 = 0.42, 0.63, 0.84, 1.0$ and $1.26$.
The pitch angle, $i_{\rm s}$, controls the strength of the torque from a spiral arm. 
Observations show that the pitch angle of the main arms ranges from $9^\circ$ to $19^\circ$~\citep[][]{Xu+2018review}.
In this study, we consider $i_{\rm s} = 13.5^\circ$, which is approximately the median value among the observed values.

Surface density $\Sigma_{\rm s} \equiv 2\rho_{\rm s} h_{\rm s}$ and $R_{\rm s}$ are key parameters in controlling radial migration around the CR radius~\citep[e.g.][]{MinchevFamaey2010}.
Based on observational estimates~\citep[][]{Siebert+2012,Eilers+2020}, we selected $\Sigma_{\rm s}/\Sigma_{\rm disk}$ values between 0.1 and 0.3. 
\citet{Mata-Chavez+2019} estimated $R_{\rm s}/R_{\rm d}=$ 1.05--1.4 using $N$-body simulations.
Therefore, we selected $R_{\rm s}/R_{\rm d}=$ 1.25 and 2 for this study.
With these settings, the distributions of maximum torque strength align closely between the SDW and DYN models, ensuring comparable radial distributions of torque strength in both models.

\section{Calculating the Frequency of Lethal GRBs}
\label{sec:evalGRB}

We provide a summary of the methodology of \citet{Spinelli+2021} to evaluate the frequency of lethal GRBs. Their approach estimates the GRB occurrence rate based on the relationship between the cosmic SFR and the frequency of SGRB or LGRBs. In this framework, the GRB frequency in different regions of the Milky Way is evaluated by correlating the local SFR with the corresponding GRB rates. The rate of lethal GRBs, $N_{\rm GRB}(R,t)$, is calculated as follows:
\begin{equation}
    N_{\rm GRB}(R,z) = \int\xi(L,z)V_{\rm MW}(z)P(d,z|R)dL,
\end{equation}
where $\xi(L,z)$ is the GRB occurrence rate as a function of luminosity $L$, $V_{\rm MW}(z)$ is the cosmological volume of the Milky Way, and $P(d,z|R)$ is the probability of a lethal GRB occurring within distance $d$ from radius $R$ in the galaxy.

\begin{figure*}
\begin{center}
\includegraphics[width=0.95\textwidth]{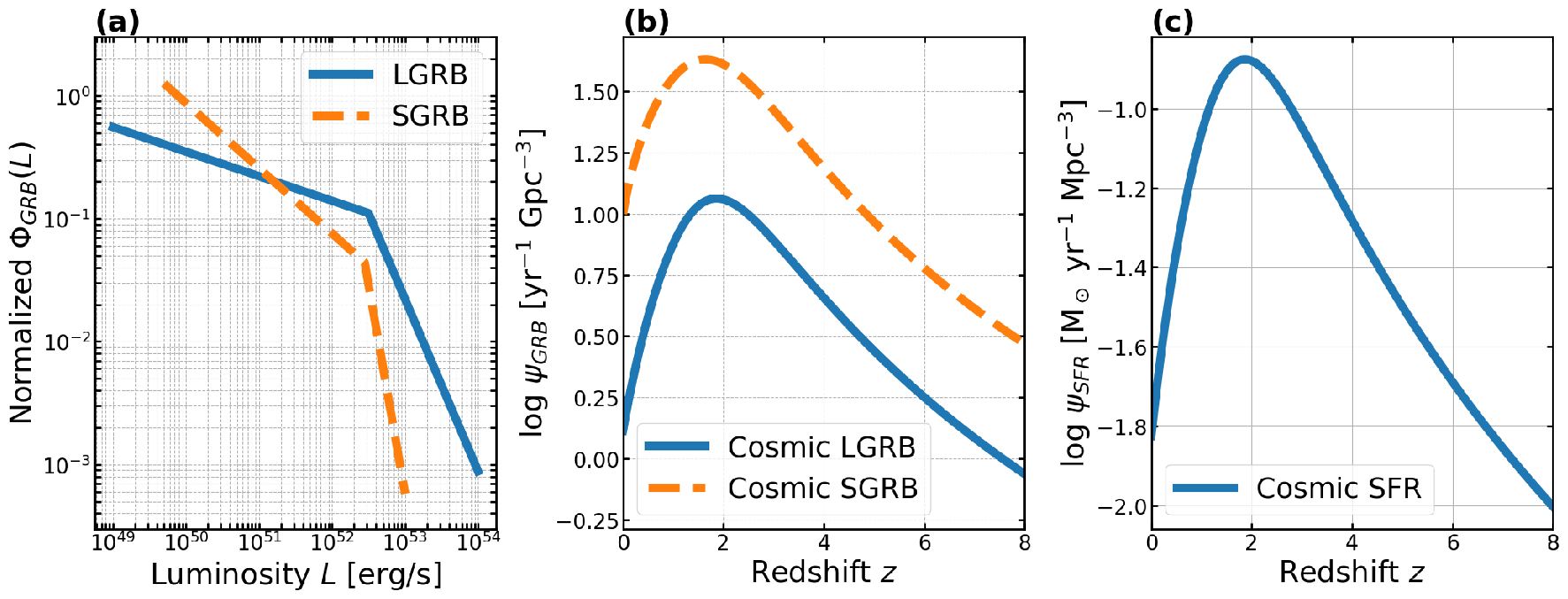}
\caption{
    Panel (a) normalized luminosity functions at $z=0$.
    Panel (b) cosmic GRB rates.
    Panel (c) cosmic SFR density.
}	
\label{fig:grb}
\end{center}
\end{figure*}

The occurrence rate $\xi(L,z)$ is defined by the normalized luminosity function $\Phi(L)$ at $z=0$ and the cosmic formation rate $\Psi(z)$.
The GRB luminosity function $\Phi(L)$ is assumed to follow a broken power law for both SGRBs and LGRBs (Fig.~\ref{fig:grb}a).
However, the rates of these two types of GRBs depend on different factors. The SGRB rate, $\Psi_{\rm SGRB}(z)$, is assumed to be a function of redshift only and does not depend on star formation rate or metallicity. In contrast, the LGRB rate, $\Psi_{\rm LGRB}(z)$, depends on both the SFR density $\Psi_{\rm SFR}(z)$ and the metallicity of the galaxy, as LGRBs are typically linked to the collapse of massive stars in low-metallicity environments \citep[][]{Virgili+2011}. The normalizations of the rates $\Psi_{\rm LGRB}(z)$ and $\Psi_{\rm SGRB}(z)$ are based on observations \citep[][]{Nakar+2006,WandermanPiran2010}. Figures~\ref{fig:grb}(b) and ~\ref{fig:grb}(c) show the redshift evolution of $\Psi_{\rm LGRB/SGRB}(z)$ and $\Psi_{\rm SFR}(z)$, respectively.

In this work, we define a ``lethal radiation hazard'' as the situation where the energy flux $F$ exceeds the critical threshold $F_c = 10^8 \, \text{erg/cm}^2$, following \citet{Thomas+2005} and \citet{Spinelli+2021}. In the case of a GRB, the energy emitted $E$ can be expressed as $E = L \cdot \tau$, where $\tau$ is the duration of the burst, set to $\tau = 20 \, \text{sec}$ for LGRBs and $\tau = 2 \, \text{sec}$ for SGRBs. The energy flux $F$ at a distance $d$ from the GRB source is then given by $F = E/(4\pi d^2)$.
Consequently, if a GRB occurs within a critical distance $d_c = \sqrt{E/(4 \pi F_c)}$, the solar system would receive an energy flux $F \geq F_c$, resulting in lethal radiation exposure.
We define the region within a distance $d < d_c$ from a position $q$ in the galaxy as ``$S_{\rm lethal}$''.

The likelihood of a GRB event $P(d,z|R)$ occurring within a specific region is refined by incorporating the stellar surface density and the specific SFR at that location, providing a detailed assessment of GRB risks across different parts of the galaxy. $P(d,z|R)$ is calculated as follows:
\begin{equation}
    P(d,z|R) = \frac{1}{M_\ast(z)}\int_{S_{\rm lethal}}\Sigma_\ast(q,z)f_{\rm sSFR}(q,z)dq,
\end{equation}
where $q$ represents the positions within the galaxy affected by the event, $\Sigma_\ast(q,z)$ is the stellar surface density at position $q$ and redshift $z$, and $f_{\rm sSFR}$ is a specific SFR correction factor. This factor adjusts the calculation based on the star formation activity relative to the average cosmic rate at the same epoch, providing a nuanced view of the risk posed by GRBs across different galactic regions. This factor is defined by 
\begin{equation}
    f_{\rm sSFR}(R,z) = \frac{{\rm sSFR}(R,z)}{{\rm sSFR}(z)},
\end{equation}
which describes the fraction of the sSFRs within the Milky Way, where ${\rm sSFR}(R,z) = \Sigma_{\rm SFR}(R,z)/\Sigma_\ast(R,z)$, relative to the specific cosmic SFR at the same epoch ${\rm sSFR}(z)=\Psi_{\rm SFR}(z)/\rho_\ast(z)$. In our study, $\Sigma_\ast(R,z)$ and $\Sigma_{\rm SFR}(R,z)$ are derived from the results of our Galactic chemical evolution model~\citep[][]{Saitoh2017,Baba+2023}.

\bibliography{ms}{}
\bibliographystyle{aasjournal}

\end{document}